\documentclass[reprint,
nofootinbib,
amssymb,
unsortedaddress,
]{revtex4-1}

\usepackage[intlimits]{amsmath}

\usepackage{graphicx}% Include figure files
\usepackage{dcolumn}% Align table columns on decimal point
\usepackage{bm}% bold math
\usepackage[english]{babel}

\usepackage{tikz}
\usetikzlibrary{arrows.meta}
\usepackage[compat=1.1.0]{tikz-feynman}

\usepackage{eufrak}
\usepackage{verbatim}

\usepackage[colorlinks=true, linkcolor = blue,
urlcolor  = blue,
citecolor = blue]{hyperref}

\begin{document}

\title{Field theory vacuum and entropic dark energy models}

\author{Michael~Maziashvili}
\email{maziashvili@iliauni.edu.ge}

\affiliation{\vspace{0.2cm} School of Natural Sciences and Medicine, Ilia State University,\\ 45 Ilia Chavchavadze Ave., Tbilisi 0179, Georgia}

% \date{\today}% It is always \today, today,
             %  but any date may be explicitly specified

\begin{abstract} 
	
 While in the standard quantization the energy spectrum of oscillator
 does not depend on its mass - for Planck length deformed quantization the energy spectrum becomes mass
 dependent. That means that the field oscillator masses will source a gravitational field through the Nullpunktsenergie as long as we follow this scheme of quantization. Admitting these masses are tangible, their gravitational effect will manifest itself even within the framework of standard field theory. We shall consider the possible gravitational implications based on this approach. If the mass scale for field oscillators is set by the inverse size of the box 	containing the field and three-momentum cutoff dictated by the black hole energy bound is exploited, one finds that the number of Fourier modes saturates the black hole entropy bound. Following certain "holographic" reasoning, one can derive various kinds of dark energy models that maybe interesting for further study.           
   
%\begin{description}

%\item[PACS numbers]

%\end{description}
\end{abstract}

%\pacs{Valid PACS appear here}% PACS, the Physics and Astronomy
                             % Classification Scheme.
%\keywords{Suggested keywords}%Use showkeys class option if keyword
                              %display desired
\maketitle
%\tableofcontents

\section{Introduction}

It is well known that the cosmological constant (CC) can not exist separately from causes and conditions tenable from a physics perspective. In quantum theory, CC is not a constant but rather it depends on the "renormalization scale" at which one probes the theory. This conclusion can be naturally reached by calculating the quantum corrections that is usually done on the basis of effective action formalism \cite{Baum:1983iwr, Hawking:1984hk, Coleman:1988tj, Adler:1989rj, Greensite:1992xw, Elizalde:1993ee, Elizalde:1993ew, Elizalde:1994av, Babic:2001vv, Guberina:2002wt, Shapiro:2003dtk, Bauer:2005rpa, Polonyi:2004ay}. (Throughout of this paper we shall adopt natural units $c=\hbar=1$.) One can recall a simple (textbook) example of Nullpunktsenergie that gives an idea of running CC \cite{Brown:1992db}

\begin{align}
&	\mathsf{\Lambda} =	\frac{\mu^{4-d}}{2}\int\frac{\mathrm{d}^{d-1}k}{(2\pi)^{d-1}} \sqrt{\mathbf{k}^2+m^2} + \Lambda_0 = \nonumber \\ & -\frac{m^4}{64\pi^2}\left[\frac{2}{4-d} +\frac{3}{2}-\gamma - \ln \frac{m^2}{4\pi \mu^2} \right] +\Lambda_0 ~.  \nonumber 
\end{align} We have used here the dimensional regularization and therefore the integral is multiplied by $\mu^{4-d}$ to ensure that Nullpunktsenergie has correct dimension. Thus $\mu$ is an arbitrary scale
included on dimensional grounds and $\gamma$ stands for the Euler-Mascheroni constant. The bare cosmological constant consists of a counter-term and the finite part. The renormalized parameter $\Lambda$ depends on $\mu$ in such a way as to make $\Lambda_0$ independent of this scale 

\begin{align}
	\mu \frac{\mathrm{d}\Lambda}{\mathrm{d}\mu} = \frac{m^2}{32\pi^2} ~. \nonumber 
\end{align} Of course, the study of the cosmological implications requires the renormalization of quantum fields in
curved space-time \cite{Bertini:2024onw} but the key point we want to stress here is that quantum theory tells us how certain quantities are running and what remains is to ascertain their values at a particular scale from experiments. For this purpose, however, one needs to specify the renormalization scale in terms of some measurable quantity. For instance, it can be identified with a characteristic energy or momentum transfer of the scattering process or with the mass of an unstable particle if quantum corrections have to do with a decay process. Besides, discussing the vacuum energy, we typically use two more scale: UV and IR. It has been argued in \cite{Taylor:1989tm, Taylor:1989ua} that the quantum correction of gravitational effective action will also depend on the IR scale that in the "large distance regime" will ensure the suppression of the cosmological constant\footnote{This result is based on the path integral formulation of Euclidean quantum gravity.}. An important qualitative step is then to set these IR, UV or renormalization scales by some cosmological parameter. It may be the age of the universe, the Hubble radius or the temperature of an appropriate cosmic fluid and so forth. Black hole (BH) physics gives us a clue about the UV/IR relationship in quantum field theory (QFT) \cite{Cohen:1998zx}. Namely one can constrain QFT by imposing BH constraints on entropy and energy. BH entropy bound implies that the physical degrees of freedom in any region scales with the radius squared, rather than cubed as might be expected \cite{Bekenstein:1980jp, tHooft:1993dmi}. Putting the field into a box with large but finite volume $l^3$ and introducing UV cut-off $\mathsf{K}$, one estimates the number of Fourier modes as $N \propto (l \mathsf{K})^3$. Therefore $N$ would scale as volume, $l^3$, if we do not impose the relation between $l$ and $\mathsf{K}$. Following this holographic reasoning that there is an upper limit set by the BH entropy to any physical degrees of freedom, one arrives at the relation

\begin{align} (l\mathsf{K})^3 \, \lesssim \,   \left(\frac{l}{l_P}\right)^2~. \nonumber  \end{align} From this relation one finds an upper bound on $\mathsf{K}$,  

\begin{align}\label{Lambda1} \mathsf{K} \,\simeq \, \frac{1}{l_{P}^{2/3} l^{1/3}} ~.\end{align}

\noindent The other UV-IR relation comes due to the limit set by forming a BH. Evaluating Nullpunktsenergie for the field with UV and IR cutoffs set by $\mathsf{K}$ and $l$, respectively, as $\mathsf{K}^4$ and demanding that associated gravitational radius $l_P^2l^3\mathsf{K}^4$ is smaller than the size of the box, $l_P^2l^3\mathsf{K}^4 \lesssim l$, one finds an upper bound on the UV cutoff

\begin{align} \label{Lambda2} \mathsf{K} \,\simeq \, \frac{1}{l_{P}^{1/2} l^{1/2}} ~.\end{align} 

\noindent These two relations, Eqs.(\ref{Lambda1}, \ref{Lambda2}), are obviously different. From the relation \eqref{Lambda2} it follows that

\[N \,\simeq\, (l \mathsf{K})^3 \,\simeq \, \left(\frac{l}{l_{P}} \right)^{3/2}  \,<\, \left(\frac{l}{l_{P}} \right)^{2}~.\] The question arises: Can one "devise" vacuum energy for the field theory that in view of the BH energy bound would yield the maximum degrees of freedom dictated by the BH entropy bound?

\section{Vacuum energy: A novel point of view}
\label{section2}

A brief summary of the previous section is that we do not have an unique picture of QFT vacuum energy but we have certain guiding ideas that lead us to one model or another. The model we want to add to the long list of already existing dark energy models emanates to some extent from observation that deformed quantum mechanics considered in \cite{Kempf:1996fz, Chang:2001kn} leads to the mass-dependent correction to the oscillator energy spectrum. Applied to the field theory, this sort of deformed quantization implies the appearance of an arbitrary length scale in QFT that is related to the masses of field oscillators \cite{Mania:2009dy, Berger:2010pj}. But mass gravitates and potentially we arrive at the result that the masses of field oscillators \cite{Fermi:1932xva, Bogolyubov:1959bfo} contribute to the vacuum energy. If we admit to consider deformed quantization as an expected prescription at certain tiny length scale of the background space, then the above mentioned result tells us that one should also ponder over the classical contribution of field oscillator masses to the vacuum energy. But what might be an emerging picture if we accept the latter idea and forget about the deformed quantization at least for a while? To address this question, let us recall that for introducing the notion of particles/quanta, the free field is usually treated as an assemblage of independent harmonic oscillators \cite{Fermi:1932xva, Bogolyubov:1959bfo}. For simplicity let us consider a free, real, massless scalar field in the finite volume $l^3$. Imposing periodic boundary conditions and using the Fourier decomposition, the Hamiltonian takes the form\footnote{The creation and annihilation operators, $\hat{a}(\mathbf{k}_n), \hat{a}^+(\mathbf{k}_n)$, that are associated to the variables $a(\mathbf{k}_n), a^*(\mathbf{k}_n)$, satisfy the commutation relation $\left[\hat{a}(\mathbf{k}_n), \hat{a}^+(\mathbf{k}_m) \right] = l^3\delta_{\mathbf{k}_n\mathbf{k}_m}$. Regarding this Hamiltonian, one may wonder about the possible role of Friedmann-Lema\^{i}tre-Robertson-Walker background space \cite{Ford:2021syk, Kolb:2023ydq}. Since this background does not affect the field oscillator masses but rather alters their frequencies (makes them time-dependent), such an approach would be unnecessary complication for our discussion.}

\begin{align}
	\label{Hamiltonian} H = \frac{1}{2l^3}   \sum\limits_{\mathbf{k}_n}  \omega_{\mathbf{k}_n} \left[ a^*(\mathbf{k}_n)a(\mathbf{k}_n) + a(\mathbf{k}_n)a^*(\mathbf{k}_n) \right] ~,
\end{align} where $\omega_{\mathbf{k}_n} \equiv |\mathbf{k}_n|$ and

\begin{align}
	\mathbf{k}_n = \frac{2\pi}{l} (n_1,\,n_2,\,n_3)~.  \nonumber 
\end{align} After introducing the real variables 

\begin{align} & Q_{\mathbf{k}_n} =\, \sqrt{\frac{1}{2l^3\mu \omega_{\mathbf{k}_n} }} \left[ a(\mathbf{k}_n) + a^*(\mathbf{k}_n)\right]~, \nonumber \\ & P_{\mathbf{k}_n} = \, i\sqrt{\frac{\mu\omega_{\mathbf{k}_n}}{2l^3}} \left[a^*(\mathbf{k}_n) - a(\mathbf{k}_n)\right] ~,   \nonumber 
\end{align} the Hamiltonian \eqref{Hamiltonian} splits into a sum of independent one-dimensional oscillators 

\begin{align} \label{oscillsum} H =   \sum\limits_{\mathbf{k}_n} \left( \frac{ P_{\mathbf{k}_n}^2}{2\mu} + \frac{\mu \omega_{\mathbf{k}_n}^2 Q_{\mathbf{k}_n}^2}{2} \right)~,\end{align} each having the mass $\mu$, which is so far chosen arbitrarily. In the framework of QFT these masses of field-oscillators \eqref{oscillsum} are usually disregarded for the quantum energy spectrum of oscillator is mass-independent. These masses, under assumption they are really existing, will provide a classical contribution to the QFT vacuum energy. Having this kind of description, at the classical level the vacuum energy of the field is represented by the unexcited field-oscillators enclosed in the box $l^3$ each of them having the mass $\mu$. One then obtains 

\begin{align}\label{mulambda}
	\rho = \mu \mathsf{K}^3 ~. 
\end{align} We can now repeat verbatim the arguments based on BH energy and entropy bounds. From the BH energy bound one obtains

\begin{align}\label{BHenergybound}
	l_P^2\mu (l\mathsf{K})^3 \, \lesssim  \, l ~ , 
\end{align} implying an upper limit on the UV cutoff

\begin{align}
	\mathsf{K} \simeq \frac{1}{l_P^{2/3}\mu^{1/3}l^{2/3}} ~ \Rightarrow ~ N \simeq (l\mathsf{K})^3 \simeq \frac{l^3}{l_P^2 l^2 \mu} ~. \nonumber 
\end{align} Thus, the particular value $\mu=l^{-1}$ ensures that BH energy bound results in the maximum value for the degrees of freedom of the field. It would be obviously wrong conceptual footing to propose that $\mu$ is merely determined by the IR cutoff. For instance it would alter the Casimir force \cite{Casimir:1948dh}. Namely, considering for simplicity a one-dimensional example \cite{Mostepanenko:1988bs} with two mirrors placed at $x=0$ and $x=d$ and the third auxiliary mirror placed at $x=\text{D}(\gg d)$, one finds for vacuum energy 

\begin{widetext}
\begin{align}
	\mathcal{E} = \left(\frac{1}{d}+\frac{\pi}{2d}\right)\sum_{j=1}^\infty j  +  \left(\frac{1}{\text{D}-d}+\frac{\pi}{2(\text{D}-d)}\right)\sum_{j=1}^\infty j =  \frac{2+\pi}{2d}\sum_{j=1}^\infty j  +  \frac{2+\pi}{2(\text{D}-d)}\sum_{j=1}^\infty j ~. \nonumber 
\end{align} The treatment of these divergent series is similar to the original approach by Casimir. Introducing the cutoff function \cite{Fierz:1960zq}, one can write 

\begin{align}&
	\frac{2+\pi}{2d}\sum_{j=1}^\infty j  \equiv \alpha  \sum_{j=1}^\infty j \, \to \, \alpha  \sum_{j=1}^\infty j \mathrm{e}^{-\alpha j L } = - \frac{\partial }{\partial L } \sum_{j=0}^\infty  \mathrm{e}^{-\alpha j L } = - \frac{\partial }{\partial L } \frac{1}{1 - \mathrm{e}^{-\alpha  L }} =  \frac{\alpha \mathrm{e}^{-\alpha  L }}{(1 - \mathrm{e}^{-\alpha  L })^2} = \nonumber \\& \frac{\alpha \mathrm{e}^{\alpha  L }}{(\mathrm{e}^{\alpha  L } - 1)^2} =  \frac{1}{\alpha L^2} - \frac{\alpha}{12} +O(\alpha^3L^2) = \frac{2d}{(2+\pi) L^2} - \frac{2+\pi}{24d} + \cdots  ~.   \nonumber 
\end{align} To remove the cutoff, we should take the limit $L\to 0$ at the end of our calculations. Therefore, terms proportional to the positive powers of $L$ can be omitted safely. In a similar manner one obtains

\begin{align}
	\frac{2+\pi}{2(\text{D}-d)}\sum_{j=1}^\infty j  \, \to \,  \frac{2(\text{D}-d)}{(2+\pi) L^2} - \frac{2+\pi}{24(\text{D}-d)} + \cdots  ~.   \nonumber 
\end{align} As a result, the Casimir force acting on the middle mirror takes the form

\begin{align}
	&	F_x = - \frac{\partial }{\partial d} \left[ \frac{2+\pi}{2d}\sum_{j=1}^\infty j  + \frac{2+\pi}{2(\text{D}-d)}\sum_{j=1}^\infty j  \right] =  \nonumber \\& - \lim\limits_{L\to 0} \left[ \frac{2}{(2+\pi) L^2} - \frac{2}{(2+\pi) L^2}  + \frac{2+\pi}{24d^2} - \frac{2+\pi}{24(\text{D}-d)^2} +\cdots  \right]  =  -\frac{2+\pi}{24d^2} + \frac{2+\pi}{24(\text{D}-d)^2} ~. \nonumber 
\end{align} \end{widetext} That is the net force on the mirror placed at $x=d$, which does not contain any more a divergent contribution. Now if we remove auxiliary mirror by taking $\text{D}\to \infty$, we arrive at the result 

\begin{align}
	F_x =  -\frac{2+\pi}{24d^2} ~. 
\end{align} In the standard case the Casimir force is $-\pi/24d^2$ and one concludes that it has now increased by the factor $(2+\pi)/\pi \approx 1.64$.  

On the other hand, if one sets the scale $\mu$ by the UV cutoff $\mathsf{K}$, then the vacuum energy density is just $\mathsf{K}^4$. To proceed our discussion in the cosmological context, one has to choose the scale $\mu$ in an appropriate way. Of course there is no unique choice but rather several natural possibilities.

\section{Age-entropic DE model}

As was mentioned above, we now feel the need for some new assumptions about $\mu$ on which to build up a dark energy model. One could follow the reasoning similar to what was put forward in \cite{Maziashvili:2006ey, Maziashvili:2007dk,Cai:2007us, Wei:2007zs} and suggest that due to energy-time uncertainty relation one can set the scale $\mu^{-1}$ by the age of the Universe

\begin{align}
	\tau = \int_0^a \frac{\mathrm{d}a}{aH} ~. \nonumber 
\end{align} The other quantity determining the DE in our approach is the degree of freedom $N$ which has to be evaluated by some holographic reasoning \cite{Fischler:1998st, Hsu:2004ri, Li:2004rb}. For this purpose one can use either particle horizon or event horizon 

\begin{align}
	d_p = a\int_0^a \frac{\mathrm{d}a}{Ha^2}\equiv \eta(a) a ~, ~~~~ d_e = a\int_a^\infty \frac{\mathrm{d}a}{Ha^2}  ~.  \nonumber 
\end{align} Here $\eta(a)$ stands for the conformal time. In view of the holographic principle \cite{Bekenstein:1980jp, tHooft:1993dmi} $N\propto (d_{p,e}/l_P)^2$. We then have for the dark energy density

\begin{align}
	\rho_{p,e} = \frac{3\alpha^2 m_P^2}{\tau d_{p, e}} ~. \nonumber 
\end{align} Here we introduced an adjustable parameter $\alpha$ to account for certain corrections that have been left out of account for the sake of simplicity\footnote{For instance the model is affected by the number of quantum fields in the universe. Also instead of Minkowski space-time one has to consider fields in the curved background and so on.}. Let us first see if this kind of DE can provide an accelerated expansion under assumption that it is a predominant component

\begin{align}\label{PDE}
	H^2  = \frac{8\pi \alpha^2}{\tau d_p} ~.  
\end{align} Looking for the solution in the form $H^{-1} = H_0^{-1} a^{\beta}$, one finds

\begin{align}\label{Phorizon}
	\beta = \frac{1+\sqrt{1+1/2\pi\alpha^2}}{2} ~.
\end{align} On the other hand 

\begin{align}
	a(t) = \left[\beta\big(H_0 t + C\big)\right]^{1/\beta} ~, \nonumber 
\end{align} and thereby a particular value of $\beta$ given by Eq.\eqref{Phorizon} can not ensure an accelerated expansion of the universe. One faces similar problem in the holographic DE model \cite{Li:2004rb}. Thus we are left with    

\begin{align}
	\rho_{e} = \frac{3\alpha^2 m_P^2}{\tau d_{e}} ~.   \nonumber 
\end{align} In this case, solving the equation similar to \eqref{PDE} gives

\begin{align}
	\beta_\pm = \frac{1\pm\sqrt{1-1/2\pi\alpha^2}}{2} ~.   \nonumber 
\end{align} Obviously $0 < 1-\beta_\pm <1$. In contrast to the holographic and agegraphic models \cite{Li:2004rb, Cai:2007us}, we have now two solutions both ensuring the accelerated expansion of the universe. It is certainly instructive to evaluate the equation of state parameter   

\begin{align}
	\omega = -\, \frac{2}{3} -\, \frac{1}{3d_eH} +\frac{1}{3\tau H} = -\, \frac{2}{3} +\frac{1}{3\tau H}  -\,\frac{\tau H \Omega}{3\alpha^2 8\pi} ~.  \nonumber 
\end{align}

\noindent It is also instructive to compare it with the equation of state parameter of agegraphic DE

\begin{align}
	\omega = -1 +\frac{2}{3\tau H} = -1 +\frac{2\sqrt{\Omega}}{3\alpha \sqrt{8\pi}} ~. \nonumber 
\end{align} Let us note that in the case of holographic DE we have 

\begin{align}
	\omega = -\frac{1}{3} - \frac{2}{3d_e H} = -\frac{1}{3} - \frac{2\sqrt{\Omega}}{3\alpha \sqrt{8\pi}}  ~. \nonumber 
\end{align}

Of course, the scale $\mu$ can be set in a number of different ways that may lead one to appealing alternative models.

\section{Some other derived models of DE}

Following the line of reasoning put forward in the previous section, one may alternatively set $\mu = H$ that gives 

\begin{align}
	\rho = \frac{3\alpha^2 m_P^2H}{ d_e} ~.  \nonumber 
\end{align} If DE dominates then this model is analogous to the Holographic DE 

\begin{align}
	-\, \frac{1}{a^2H} = \frac{\mathrm{d}}{\mathrm{d}a} \frac{8\pi\alpha^2}{aH} ~,  \nonumber 
\end{align} The equation of state parameter has the form

\begin{align}
	\omega = -\frac{2}{3} - \frac{1}{3d_e H} - \frac{\dot{H}}{3H^2} ~.  \nonumber 
\end{align}

The other class of DE models maybe derived immediately from Eq.\eqref{mulambda} by plugging in it $\mu = \mathsf{K}^\beta l^{\beta-1}$ and setting the scale $l$ either by $\tau$ or $H^{-1}$. In the former case we obtain the generalization of models discussed in the framework of different approaches in \cite{Zee:1983jg, Zee:2004um, Maziashvili:2007vu} 

\begin{align}
	\rho = \frac{\mathsf{K}^{\beta+3}}{\tau^{1-\beta}} ~.    \nonumber 
\end{align} In the latter case one would obtain the DE density 

\begin{align}
	\rho = \mathsf{K}^{\beta+3} H^{1-\beta} ~,   \nonumber 
\end{align} that generalizes the model \cite{Zhitnitsky:2011tr} in which $\mathsf{K}$ is set by the QCD scale.   

\section{Concluding remarks}

In standard quantum theory it makes no difference which scale is used for setting the masses of field oscillators since the energy spectrum of quantum oscillator is mass independent and it is a common practice to use the size of a box (enclosing the field) to set the masses of these oscillators \cite{BornJordan}. It is curious, however, to remark that in Planck length deformed quantization scheme \cite{Kempf:1996fz, Chang:2001kn} the energy spectrum of harmonic oscillator becomes mass dependent \cite{Mania:2009dy} that means that the Nullpunktsenergie in QFT will depend on the masses of field oscillators. In other words, these masses will source a gravitational field via the Nullpunktsenergie. But if these masses are tangible (realistic), they will gravitate even in the standard quantization scheme just in the classical manner. Thus, the essential idea is to admit that field oscillators gravitate because of their masses irrespective to the way the mass scale is introduced. Considering the possible gravitational implications of this mass scale, one finds a curious fact that when this scale is set by the size of the box (containing the field) and the three-momentum cutoff dictated by the black hole energy bound is exploited - the number of Fourier modes saturates the black hole entropy bound. 

Our discussion was focused on boson field. What about the fermion fields? As it is well known, fermion contribution to the Nullpunktsenergie is negative (in contrast to the boson one) because commutators are replaced by anti-commutators in the normal ordering prescription. This result goes beyond the zero-point level providing the basis to the remarkable feature of the supersymmetric theory that its vacuum energy equals zero \cite{Zumino:1974bg}. However, the vacuum energy introduced in section \ref{section2} has nothing to do with the quantization prescription and works for the fermion field in the same manner. It is always positive definite.

Loosely speaking, the scale $\mu$ introduced in section \ref{section2} is somewhat similar to the renormalization scale that requires further specification for obtaining definite predictions. In the fixed-order perturbative calculations in QFT, the renormalization scale is usually assumed to be a typical momentum transfer of the process. For specifying this kind of scale in the cosmological context one usually thinks of some cosmological parameter such as the age of the Universe \cite{Maziashvili:2006ey, Maziashvili:2007dk, Cai:2007us, Wei:2007zs}; the particle horizon \cite{Fischler:1998st}; Hubble parameter \cite{Hsu:2004ri}; the future event horizon \cite{Li:2004rb}; the Ricci scalar curvature \cite{Gao:2007ep}; the combination of $H^2$ and $\dot{H}$ \cite{Granda:2008dk} and so forth. Besides the scale $\mu$, the vacuum energy density depends on the number of oscillators that is proportional to $\mathsf{K}^3$, see Eq.\eqref{mulambda}. In general, one may think of $\mu$ as a certain combination of UV and IR scales. For instance one can take $\mu = \mathsf{K}^\beta l^{\beta-1}\Rightarrow\rho = \mathsf{K}^{\beta+3} l^{\beta-1}$. If we take the UV scale at Planck energy, then one needs to set $\beta=-1$ and take $l\simeq H^{-1}$ for obtaining the correct value of the present DE density \cite{Padmanabhan:2004qc}. The other possibility is to take $\mathsf{K}=\mathsf{K}_{QCD}, \beta =0$ \cite{Maziashvili:2007vu} \cite{Zhitnitsky:2011tr}. However, for controlling the upper bound on the UV scale, one may find it reasonable to impose the black hole energy bound similar to Eq.\eqref{BHenergybound} that leads to the UV-IR relation

\begin{align}
	\mathsf{K} \simeq \frac{1}{(l_P^2l^{1+\beta})^{1/(3+\beta)}} ~. \nonumber 
\end{align}

Since the discovery of present accelerated expansion of the Universe \cite{SupernovaSearchTeam:1998fmf, SupernovaCosmologyProject:1998vns} a huge number of DE models have been suggested. The present situation, however, is quite hard for many of these models since DESI results (under the assumption of CPL parametrization) \cite{DESI:2024mwx, DESI:2025zgx} indicate that at present $\omega > -1$ while in the resent past $\omega$ was less than $-1$. If these results are confirmed, many of the DE models will be ruled out. The above suggested DE models may be interesting to address this specific issue. Let us note that a relatively general parametrization of DE has been used recently for analyzing DE dynamics \cite{Lee:2025pzo}.

One more interesting idea in defining the DE density is related to the effect of self gravitation \cite{Arnowitt:1962hi, Arnowitt:1960zz, Misner:1963zz, Duff:1973zz, Duff:1973ji, Tsamis:2011ep}. Consider a mass $M$ distributed in a sphere of radius $l$. In view of Newtonian gravity the total energy takes the form

\begin{align}
	\mathfrak{M} \,=\,	M - \frac{3l_P^2M^2}{5l} ~. \nonumber 
\end{align} But in view of general relativity it is the total energy that interacts gravitationally telling us that the above expression should be replaced by       

\begin{align}
	\mathfrak{M} \,=\,	M - \frac{3l_P^2\mathfrak{M}^2}{5l} ~ \Rightarrow ~ \mathfrak{M} = \frac{5l}{6l_P^2}\left[ \sqrt{1+ \frac{12 l_P^2 M}{5l}}  - 1\right] ~.   \nonumber 
\end{align} Thus for the energy density $\rho = 3\mathfrak{M}/4\pi l_3$ one finds

\begin{align}
	\rho = \frac{5}{8\pi l^2 l_P^2} \left[ \sqrt{1+ \mu l_P^2 l^2 \mathsf{K}^3}  - 1\right] ~, \nonumber 
\end{align} where we have absorbed a numerical factor in $\mu$. Identifying $l\simeq H^{-1}$, the factor $(ll_P)^{-2} \simeq H^2/l_P^2$ gives the correct value of the present DE density \cite{Padmanabhan:2004qc}.

\begin{acknowledgments} 
Author is indebted to Nishant Agarwal and Tina Kahniashvili for useful comments. The work was supported in part through funds provided by Shota Rustaveli National Science Foundation of Georgia under Grant No. FR-24-901 and by Ilia State University under the Institutional Development Program.
\end{acknowledgments}

\end{document}